# A Sustainable Photocatalytic Pathway for Concurrent Hydrogen and Value-Added Chemical Production Utilizing Microalgae as Bio-Scavenger in Water

Ho Truong Nam Hai[a,b], Augusto Ducati Luchessi[c] and Kaveh Edalati[a,b,d],*

[a] WPI, International Institute for Carbon Neutral Energy Research (WPI-I2CNER), Kyushu University, Fukuoka 819-0395, Japan

[b] Department of Automotive Science, Graduate School of Integrated Frontier Sciences, Kyushu University, Fukuoka 819-0395, Japan

[c] Laboratório de Biotecnologia BraPhyto, Faculdade de Ciências Aplicadas (FCA), Universidade Estadual de Campinas (UNICAMP), 13484-350, Limeira, SP, Brazil

[d] Mitsui Chemicals, Inc. - Carbon Neutral Research Center (MCI-CNRC), Kyushu University, Fukuoka, 819-0395, Japan

*Corresponding author (E-mail: kaveh.edalati@kyudai.jp; Phone: +81 92 802 6744)

icroalgae are an abundant bioorganic material source and play a significant role in life on Earth by conducting photosynthesis for carbon dioxide ($CO_2$) capture and its conversion to oxygen ($O_2$). In this study, a combination of microalgae as a negative-$CO_2$-emitting sacrificial agent with the traditional photocatalytic water-splitting process using brookite $TiO_2$, as a model photocatalyst, is introduced as a new strategy to maximize green hydrogen ($H_2$) production while converting microalgae to valuable products, like methane ($CH_4$) and carbon monoxide (CO). The process, under optimal conditions, produces up to 0.990 mmol/g.h of $H_2$ without cocatalyst addition and 3.200 mmol/g.h with platinum (Pt) cocatalyst, which is 13 times higher than the production rate without microalgae. The strategy of using microalgae in photocatalysis has high potential in green $H_2$ production, as it not only eliminates valuable hole sacrificial agents, like alcohol, but also produces other useful compounds, like $CH_4$ and CO. Moreover, this sustainable process contributes to $CO_2$ capture and conversion during microalgae cultivation.

**Keywords:** photosynthetic organisms, biogas, biomass, methanation, photoreforming

## 1. Introduction

Energy plays a crucial role in driving the economy and society, and ensuring sustainable development for every nation. However, energy shortages are increasing due to dwindling supplies of fossil fuels and oil reserves [1]. In addition, traditional fuel sources such as coal, oil and liquefied natural gas, among others, contribute significantly to environmental pollution and climate change when burned [2,3]. Therefore, enhancing the production of alternative fuels like hydrogen ($H_2$), biodiesel and biomethane to replace fossil fuels exhibits a potential and sustainable option in the near future. Currently, the main industrial method for producing $H_2$ is steam methane reforming [4,5]. This method relies on natural gas derived from fossil hydrocarbon sources. During this process, a large amount of $CO_2$ is emitted, which is a major factor causing the greenhouse effect [2] and global warming [3]. To minimize the environmental problems of $H_2$ production, using non-emission sources is a necessary option. Among them, photocatalytic water splitting is a feasible and promising method as it only uses solar radiation and photocatalysts with suitable band structures. $TiO_2$ crystalline phases, such as rutile, anatase and brookite, are the most popular catalysts for photocatalysis [6–9]. While a rutile-anatase mixture is typically considered a benchmark photocatalyst, various studies have also reported high activity of brookite for this application [10–13].

In the photocatalyst process, photogenerated electrons participate in the reduction reactions to generate $H_2$, whereas photogenerated holes participate in the water oxidation process [14]. However, the oxidation of water to oxygen ($O_2$) requires the hole to have a strong oxidation energy for the 4-electron pathway ($2H_2O \rightarrow O_2 + 4H^+ + 4e^-$) [15]. However, in most cases, holes cannot oxidize water to produce $O_2$. These holes gather on the surface of the catalyst, reducing efficiency over long-term use and sometimes oxidizing the catalyst. To eliminate this problem, hole sacrificial agents, like alcohol, are typically used during photocatalysis. Since hole sacrificial agents are valuable and usually more expensive than hydrogen, there are significant attempts to achieve overall water splitting or use organic waste as sacrificial agents [16–19]. The photocatalytic process in which waste of plastics [17], organic compounds [20–22], amino acids [23] or antibiotics [24,25] is used as sacrificial agents is known as photoreforming. In photoreforming, the use of valuable sacrificial agents is not only eliminated, but waste itself is converted by holes to value-added organic compounds. Therefore, the photoreforming process is considered a clean pathway in energy production and environmental issues. Photocatalytic photoreforming offers distinct advantages, including direct solar energy utilization, mild operating conditions, rapid reaction kinetics, and the ability to simultaneously produce hydrogen and valorize organic waste, making it competitive with electrolysis, fermentation, and photobiological routes. However, the current global policy in managing and recycling plastics and other organic waste sources, which emit $CO_2$ during their production, limits the future expansion of the photoreforming process to produce hydrogen. There is currently a high demand to explore sacrificial agents that are abundant and do not release $CO_2$ during their production.

The application of photosynthetic organisms as a sacrificial agent can be an interesting option for the photocatalytic process to make the process fully environmentally friendly. Among these, microalgae are known as a massive source of biomass [26]. Microalgae are unicellular organisms that exist in most aquatic environments on the planet [27]. They can perform photosynthesis, significantly contributing to $CO_2$ capture and conversion, producing about half of the $O_2$ on the planet [26,28]. In addition, microalgae can survive and thrive in diverse environmental conditions, from normal to extreme conditions in terms of temperature, humidity, pH and light intensity, among others [27]. The main components of most microalgae biomass include lipids, proteins and carbohydrates [29]. With their ease of cultivation and high biomass productivity [26], the authors consider microalgae as a promising sustainable resource for photocatalysis. So far, no study has elucidated the process of microalgae photoconversion through the photocatalytic method and its utilization as a hole scavenger.

In this study, we report the first photocatalytic process for simultaneous hydrogen production and microalgae photoconversion. Brookite, originally introduced as a photocatalyst in 1985 [30], is chosen as a model catalyst in the current investigation because an experiment of photocatalytic water splitting, as described in Fig. S1, and an earlier study demonstrated its superiority in the photocatalysis and photoreforming processes over the other two forms of $TiO_2$, anatase and rutile [10]. This study shows, for the first time, that microalgae offer advantages over other sacrificial agents when used in a two-stage system. In the first stage, microalgae cultivation enables the uptake of atmospheric $CO_2$. After harvesting, and in the second step, microalgae are used in a photocatalytic process, thereby enhancing $H_2$ production efficiency and contributing to the production of CO and $CH_4$ as two value-added chemicals.

## 2. Materials and methods

### 2.1. Microalgae cultivation

The wild-type microalga *Chlamydomonas reinhardtii* CC-124 strain was obtained from the Chlamydomonas Resource Center (CRC) in St. Paul, MN, USA. All growth stages were conducted under constant illumination of 48 μmol photons $m^{-2}$ $s^{-1}$. The microalgae were kept on tris-acetate-phosphate (TAP) medium with agar in petri dishes at 298 K. The recipe for the TAP medium was acquired from the CRC website as originally defined by Gorman and Levine in 1965 [31]. To obtain microalgal biomass, cultures were grown in Erlenmeyer flasks containing TAP medium at 298 K with agitation at 150 rpm on a rotary shaker, a setup that allowed for gas exchange without forced aeration. The cultivation was scaled up stepwise before transferring 110 mL of the culture to a 15 L photobioreactor containing TAP medium. Agitation in the photobioreactor was achieved using a magnetic stir bar and sparging with atmospheric air, which was filtered using a 220 nm membrane. Following cultivation, the microalgae were harvested using centrifugation at 2.00 g for 10 minutes at 298 K. The collected cell pellet was then dehydrated by lyophilization, and the resulting biomass was stored at -253 K until further usage for catalysis. It should be noted that hydration and lyophilization were used to ensure the reproducibility of the composition of cultivated and harvested microalgae.

## 2.2. Reagents and materials

Brookite with 99.99% purity was purchased from Kojundo, Japan. Sodium hydroxide (NaOH) and hydrogen hexachloroplatinate (IV) hexahydrate ($H_2PtCl_6 \cdot 6H_2O$) were supplied from Fujifilm, Japan, and Goodfellow, UK, respectively.

## 2.3. Characterization

The characteristics of brookite and microalgae were analyzed using various techniques. The crystal structure of brookite was examined using X-ray diffraction (XRD) equipped with a Copper K$\alpha$ radiation source and by the Raman spectroscopic method using a 532 nm laser excitation. The chemical composition of the microalgae was determined using Fourier transform infrared spectroscopy (FTIR) having a single-reflection attenuated total reflectance (ATR). The morphological features of both brookite and microalgae were checked utilizing a scanning electron microscope (SEM) operated at 15 keV and 5 keV, respectively. The nanostructure of brookite was further investigated using a transmission electron microscope (TEM) at 200 keV, through bright-field (BF) and dark-field (DF) micrographs, selected area electron diffraction (SAED) and high-resolution (HR) micrographs. X-ray photoelectron spectroscopy (XPS) using aluminum K$\alpha$ was employed to analyze the oxidation states of titanium and the reduction states of oxygen in brookite. To determine the optical properties, UV–vis spectroscopy was conducted in the wavelength range of 200 nm to 800 nm. Finally, the recombination behavior of photo-induced charge carriers and the characteristics of oxygen vacancies in brookite were identified through photoluminescence (PL) spectroscopy (325 nm laser excitation) and electron spin resonance (ESR) spectroscopy (9.4688 GHz microwave).

**2.4. Catalysis**

Photocatalytic experiments using microalgae were conducted in batch mode under various conditions, including variations in solvent type (deionized water and NaOH), solvent concentration (NaOH at 1 M, 5 M and 10 M), and the presence or absence of a platinum cocatalyst. These experiments aimed to find the optimal conditions for hydrogen generation and to gain insight into the role of microalgae in the reaction. In all conditions, the mass of brookite and microalgae was fixed at 50 mg each, maintaining a 1:1 mass ratio. The materials were placed in a 160 mL quartz photoreactor, followed by the addition of 27 mL of solvent. In certain experiments, 0.25 mL of $H_2PtCl_6 \cdot 6H_2O$ (0.01 M), corresponding to 1 wt% of platinum loading, was added as a source of cocatalyst. The photoreactor was sealed with a septum cap and purged for 10 minutes with argon gas to ensure an anaerobic condition. The photocatalytic system was established by placing the photoreactor in a cooling system and on continuous stirring equipment. A 300 W Xe lamp (1.3 W/cm$^2$) with a full wavelength range and no cut-off filter, positioned approximately 10 cm above the photoreactor, served as the light source. Gas sampling was performed using a syringe, collecting two 0.5 mL aliquots at the same time. One aliquot was injected into a gas chromatograph (GC) having a thermal conductivity detector (TCD) for $H_2$ analysis, while the other was analyzed using GC connected a methanizer and equipped with a flame ionization detector (FID) to determine the concentrations of $CH_4$, $CO$ and $CO_2$. The TCD detection limit was 1 μmol, and the total measurement error was under 10%.

After the photoreforming reaction, the liquid phase was filtered using a 125 mm diameter filter paper (Code: 1-1770-15, 125 mm diameter, 11 μm pore size). A 20 mL of the filtrate solution was transferred into a separatory funnel along with 5 mL hexane, 2 mL saturated KCl solution and 2 mL deionized water for liquid–liquid extraction over 30 minutes. The supernatant organic phase was collected, and the extraction was repeated once more to ensure complete separation of organic compounds from the aqueous phase. After extraction, the sample was dried under argon gas to concentrate the extract. To identify possible organic compounds generated during the reaction, 1 μL of the concentrated sample was directly injected into a gas chromatography–mass spectrometry (GC–MS) device. The GC was equipped with a CP-Sil 8CB column (split 50:1), and the temperature programming was set as follows: keep at 313 K for 2 minutes, ramp at 4 K/min to 513 K, and hold for 2 minutes. The MS was operated with an ion source temperature of 473 K, an interface temperature of 523 K and a mass scan range of $m/z$ = 50–700, following previously reported methods [32,33].

## 3. Results

Fig. 1a shows XRD spectra to characterize the crystal structure of brookite. It can be observed that brookite has an orthorhombic phase (Pcab group, $a$ = 5.458 Å, $b$ = 9.060 Å, $c$ = 5.144 Å, $\alpha = \beta = \gamma = 90°$). Fig. 1b depicts the symmetrical characteristics of brookite through Raman analysis. It is observed that four different vibrational modes, including $A_{1g}$ (134, 154, 198, 248 and 640 cm$^{-1}$), $B_{1g}$ (215, 324, 415 and 505 cm$^{-1}$), $B_{2g}$ (370, 398, 466 and 588 cm$^{-1}$) and $B_{3g}$ (288 and 549 cm$^{-1}$), are similar to the characteristics of brookite in previous studies [25,34]. Additionally, the absence of peaks at 442 cm$^{-1}$ and 516 cm$^{-1}$, that are characteristic of the Raman spectra of rutile and anatase, confirms that the material is completely free from other phases [25]. Fig. 1c depicts the UV-vis absorbance spectrum of brookite, showing that the material strongly absorbs light only in the UV range. Based on the Kubelka-Munk theory, the indirect bandgap of brookite is estimated to be around 3.0 eV, as depicted in Fig. 1d, which is consistent with earlier publications on the same brookite powder [10,35]. Since the conduction band top for this powder was reported as +2.0 eV vs. NHE (normal hydrogen electrode) [35], the conduction band bottom position of brookite is estimated as -1.0 eV vs. NHE. Since the position of the conduction band is more negative than the water reduction potential ($H_2O/H_2$: 0 eV) [36,37], the current brookite powder satisfies the thermodynamic requirements for $H_2$ production. The XPS spectrum (Fig. 1e) confirms the full oxidation state of Ti 2p in brookite. The ESR spectrum of brookite is described in Fig. 1f, which identifies unpaired electrons. The appearance of a peak at $g$ = 2.008 confirms the partial existence of oxygen vacancies in the crystal lattice. Fig. 1g represents the PL spectra to determine the recombination rate of photo-induced charge carriers in brookite. A peak appears in the 550–620 nm range, and the peak intensity reaches about 200 cps, indicating that charge recombination is not significant in brookite.

Fig. 1h describes the chemical structure characteristics of microalgae through FTIR analysis. It is seen that the composition of *Chlamydomonas reinhardtii* microalgae is characterized by various groups, including lipids, proteins and carbohydrates. Among these, the lipid group is identified based on the C–H stretching vibrations of fatty acids (2873, 2923 and 2962 cm$^{-1}$) and the C=O stretching of phospholipids (1739 cm$^{-1}$) [38–40]. The protein group is characterized based on N–H stretching (3000–3600 cm$^{-1}$), C=O stretching (1643 cm$^{-1}$), N–H bending vibrations of amide (1542 cm$^{-1}$) and P=O antisymmetric stretching vibrations of nucleic acids, phosphoryl group, or phosphorylated proteins (1242 cm$^{-1}$) [38–40]. Finally, the carbohydrate group is illustrated based on O–H stretching (3000–3600 cm$^{-1}$), C–H bending (1410 and 1450 cm$^{-1}$) and C-O-C stretching (1080 and 1154 cm$^{-1}$) vibrations of polysaccharides [38–40]. The results show that the cultivated microalgae are fully compatible with previous studies [38–40].

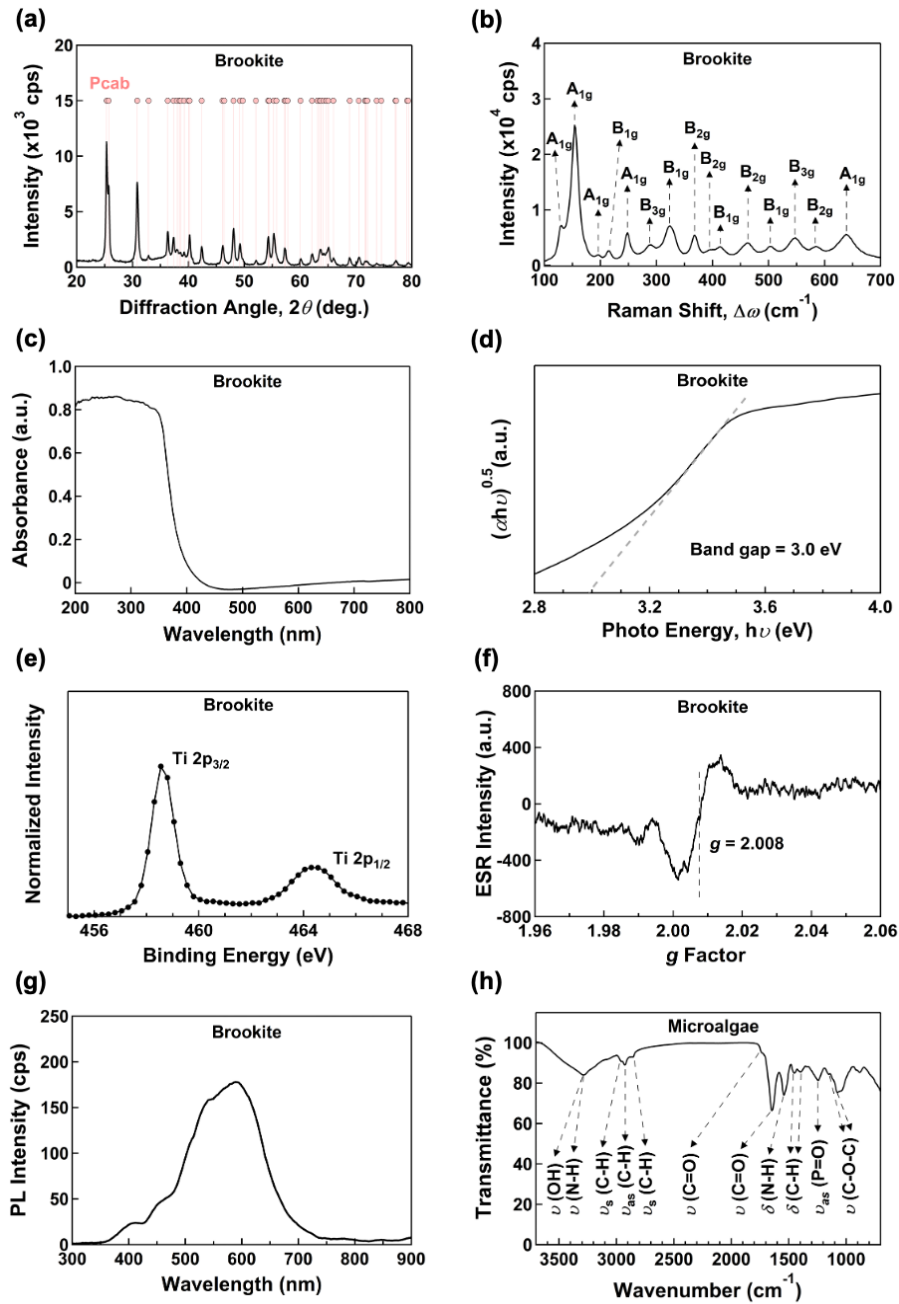

Fig. 1. Structural and optical properties of brookite and chemical properties of microalgae. (a) XRD profile, (b) Raman spectrum with different vibration modes, (c) UV-Vis absorbance spectrum, (d) Kulbelka-Munk analysis for indirect bandgap calculation ($\alpha$: absorption coefficient: h: Planck's constant, $v$: light frequency), (e) XPS Ti 2p spectrum, (f) ESR spectrum, and (g) PL spectrum of brookite. (h) FTIR spectrum with various chemical groups vibration of microalgae.

Fig. 2 shows the SEM images for (a, b) brookite and (c, d) microalgae. It is observed that brookite has a uniform shape, and its particle size ranges from 3 to 5 μm. Meanwhile, microalgae clusters show heterogeneity in shape and size, with the length of each cluster ranging from 20 μm to 250 μm. The heterogeneous morphology of microalgae clusters may also be a factor affecting their interactions during photocatalysis.

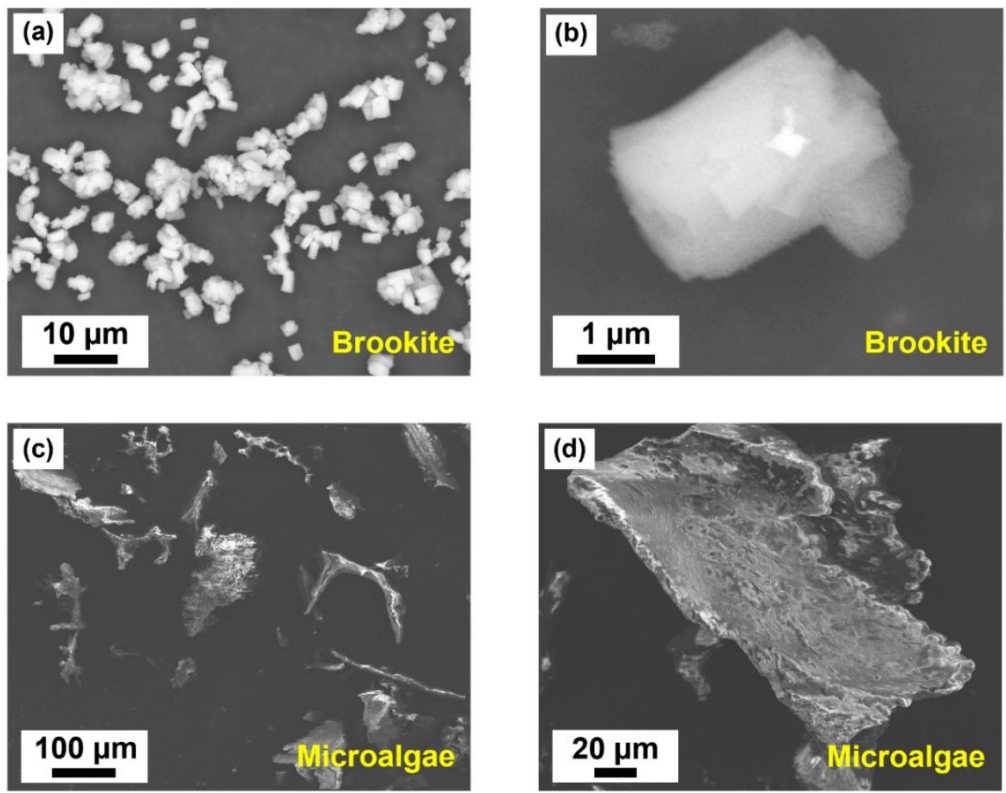

Fig. 2. Morphological characteristics of brookite particles and microalgae clusters at microscale. SEM image of (a, b) brookite particles and (c, d) microalgae clusters at different magnifications.

Fig. 3 illustrates the microstructure and nanostructure of brookite. The TEM (a) BF and (b) DF images show the presence of numerous nanocrystals within each particle. In addition, Fig. 3c shows the SAED pattern with concentric circles corresponding to the crystal planes in the lattice structure of brookite (the diameters of circles were calculated by considering $d$-spacings of atomic planes and the camera length of TEM). Fig. 3d depicts the HR image and the fast Fourier transform (FFT), which also confirms that brookite has an orthorhombic phase, as expected from the presented data in the literature [10–13].

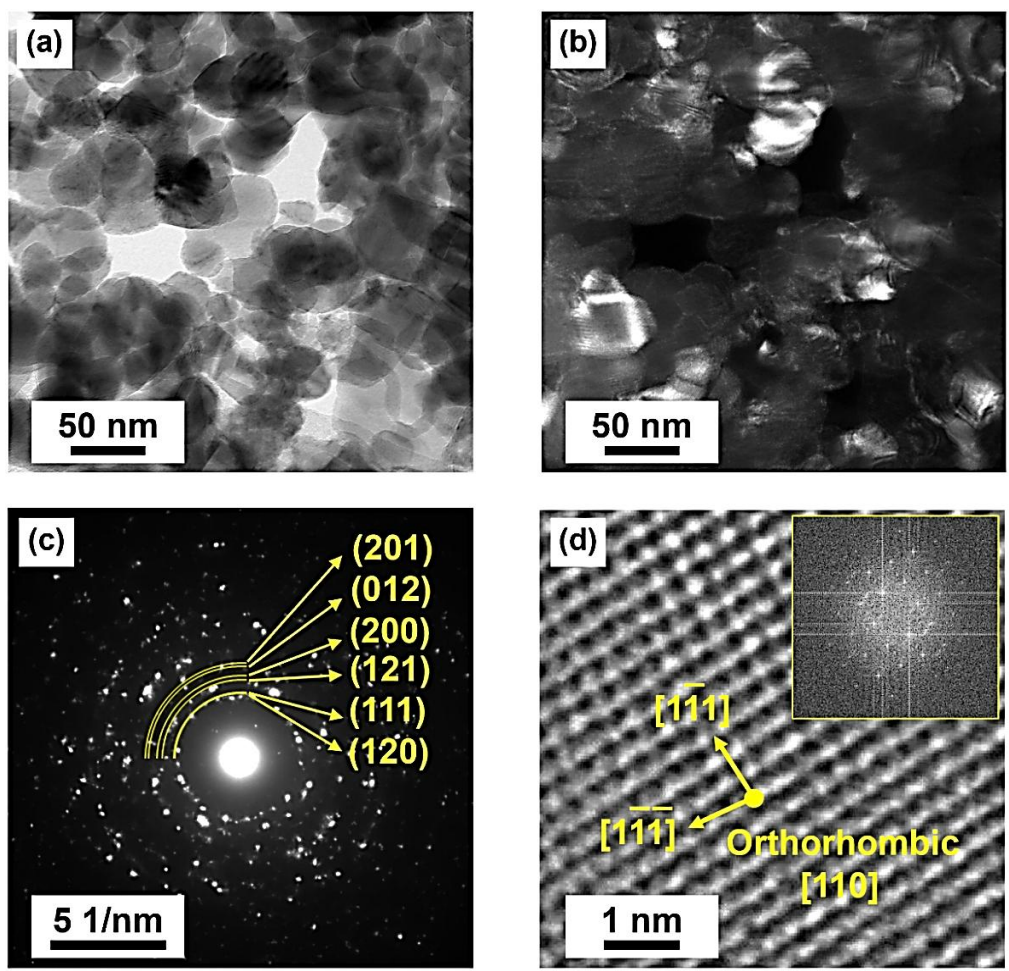

Fig. 3. Orthorhombic brookite with nanosized crysals. TEM (a) BF and (b) DF micrographs, (c) SAED analysis, and (d) HR micrograph with FFT analysis

Fig. 4a shows $H_2$ production from photocatalysis using brookite without and with microalgae addition under various conditions within 3 h of irradiation. It can be seen that the photocatalytic process of brookite in water without the addition of any other agents only produces a small amount of $H_2$, reaching 0.011 mmol/g·h. When increasing the NaOH concentration in the solution and enhancing alkalinity, brookite exhibits a higher $H_2$ production efficiency compared to neutral pH conditions. Specifically at 10 M NaOH, $H_2$ concentration reaches 0.078 mmol/g.h. This change can be explained by the aggregation of brookite particles, which occurs at pH values close to the point of zero charge ($pH_{pzc}$). In an aqueous environment with a neutral pH near the $pH_{pzc}$ value, the size of brookite particles may become significantly larger compared to that under strongly alkaline conditions far from the $pH_{pzc}$. The large particle size can affect photon absorption efficiency and consequently reduce the formation of the photogenerated electron [41]. With the presence of microalgae in the photocatalytic process, the $H_2$ production efficiency increases. This increase is significantly affected when increasing the NaOH concentration in the solution. A strongly alkaline environment accelerates the hydrolysis of microalgae into components with simpler structures, thereby enhancing the oxidation process by photogenerated holes, reducing charge recombination, and facilitating the participation of electrons in the water reduction reaction. In detail, $H_2$ production after 3 h from the microalgae photoreforming using brookite in water, 1 M NaOH, 5 M NaOH and 10 M NaOH reached 0.022, 0.032, 0.263 and 0.990 mmol/g.h, respectively. This indicates that in a solution with 10 M NaOH, the addition of microalgae increases the $H_2$ evolution rate 13 times. Adding $H_2PtCl_6 \cdot 6H_2O$ to the reaction mixture as a source of 1wt% photodeposited platinum further improves the process efficiency and $H_2$ production rate. The observations show that the surface of brookite transforms to black in the first minutes of irradiation due to the photodeposition of platinum under UV irradiation. However, in the absence of microalgae, the reaction occurs quickly at the beginning, but the rate gradually decreases over time due to a lack of a hole scavenger. The $H_2$ production rate reaches 0.247 mmol/g.h in the presence of brookite and platinum and without microalgae addition, which is surprisingly 4 times less than the one with microalgae addition and without the platinum addition.

Fig. 4b shows the concentrations of products from photocatalysis by brookite and microalgae addition under the optimal conditions (10 M NaOH and platinum addition). The $H_2$ concentration reaches 3.200 mmol/g.h, confirming 13 times increase compared to similar conditions without the microalgae addition. Additionally, microalgae usage results in the evolution of 0.030 mmol/g.h of $CH_4$ and 0.133 mmol/g.h of CO. Moreover, a negligible amount of $CO_2$ is also detected despite the high concentration of NaOH in the solution, possibly due to atmospheric $CO_2$. These results indicate that the microalgae process not only enhances $H_2$ production from water splitting but also contributes to $CH_4$ and CO formation.

To assess the role of microalgae hydrolysis in the photocatalytic reaction, Fig. 4c shows the degradation of microalgae under irradiation conditions with 10 M NaOH addition without any photocatalyst. It can be observed that the decomposition of microalgae only produces a very small amount of $H_2$, reaching 0.004 mmol/g.h. Meanwhile, this decomposition also produces 0.003 mmol/g.h of $CH_4$ and 0.033 mmol/g.h of CO, which are lower than those produced by photocatalysis. From the obtained results, it can be seen that the hydrolysis of microalgae contributes only a minor fraction to the concentrations of gaseous products, including $H_2$, CO, and $CH_4$. In comparison, the majority of these products are generated by photocatalytic reforming of microalgae using brookite in the presence of Pt and 10 M NaOH. Moreover, a small amount of detected $CO_2$ shows no correlation with time, similar to the one detected during photocatalysis in Fig. 4b.

Fig. 4d shows the performance of the microalgae photoreforming process after two cycles. In the first cycle, microalgae are consumed as a sacrificial agent. To evaluate the role of microalgae in photocatalysis, microalgae were not added in the second cycle, while the same conditions remained. Consequently, lower concentrations of $H_2$, CO, and $CH_4$ were detected in the second cycle due to the limited amount of remaining microalgae. These results indicate that microalgae contribute to the photocatalysis of water as a sacrificial agent.

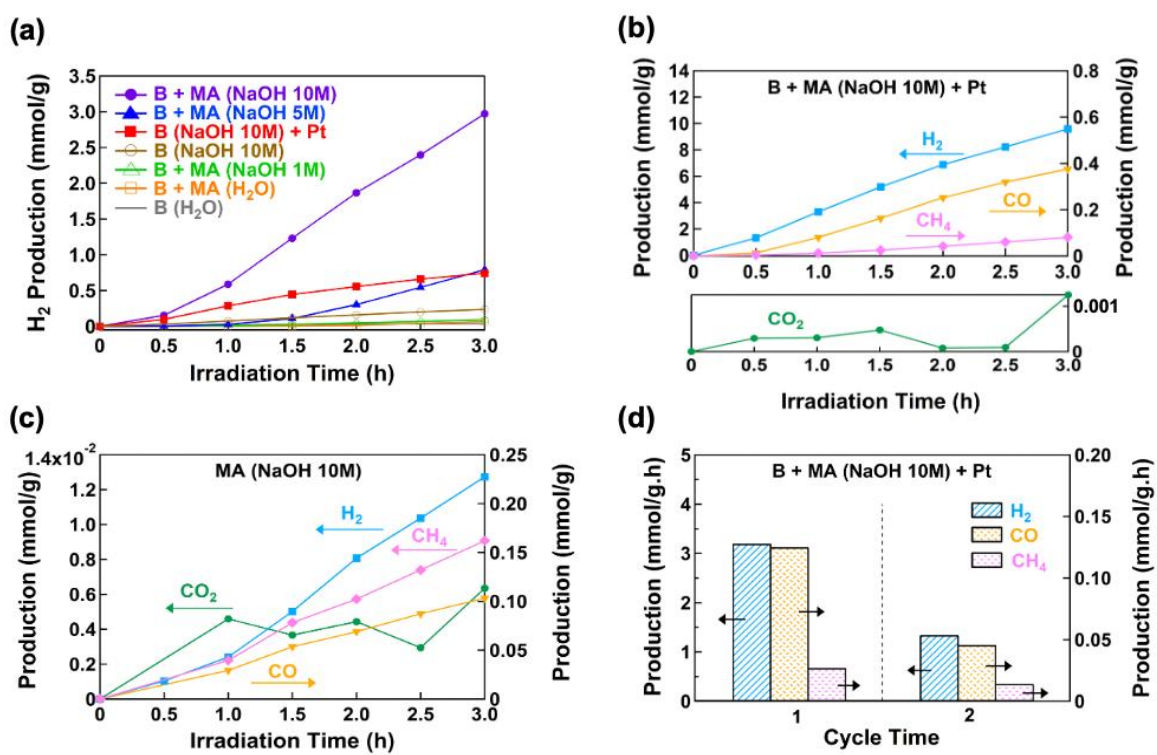

Fig. 4. Boosted catalytic formation of hydrogen and generation of methane and carbon monoxide by using microalgae as a hole sacrificial agent. (a) $H_2$ evolution from photocatalysis reactions under various conditions. (b) $H_2$, CO, $CH_4$ and $CO_2$ production by using brookite and Pt cocatalyst in the presence of 10 M NaOH and microalgae. (c) $H_2$, CO, $CH_4$ and $CO_2$ production from microalgae in 10 M NaOH without brookite addition. (d) photocatalytic production rates for two cycles by using brookite and Pt cocatalyst in the presence of 10 M NaOH and microalgae. B and MA indicate brookite and microalgae, respectively.

To better understand the pathway of microalgae transformation in photocatalytic reactions, as well as to identify possible organic products in the liquid phase after the reaction, GC–MS analysis was conducted. Fig. 5 shows the GC–MS chromatograms of hexane solvent, the extracted liquid phase from microalgae degradation (10 M NaOH) by irradiation in hexane, and the extracted liquid phase from photocatalysis by using brookite, platinum, microalgae and 10 M NaOH in hexane. It is observed that the chromatogram of microalgae in hexane is different from the other two chromatograms, as there are three peak positions at retention times of 19.315, 35.090 and 51.455. Although the authors were not successful in the identification of these three peaks using the software equipped on the GC-MS machine, they should be intermediate organic compounds formed from either lipid, carbohydrate and protein, which were detected in FT-IR [42-48].

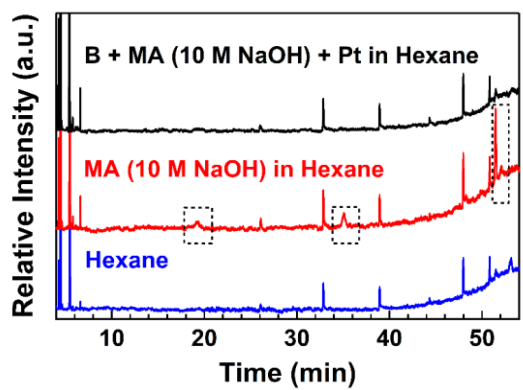

Fig. 5. Transformation of microalgae to other organic molecules by irradiation in NaOH and their disappearance by photocatalysis. GC–MS chromatograms of hexane, extracted liquid phase of 3 h irradiated microalgae and 10 M NaOH in hexane, and extracted liquid phase of 3 h photocatalysis using brookite, microalgae, Pt and 10 M NaOH in hexane. B and MA indicate brookite and microalgae, respectively.

Fig. 6 illustrates the stability of brookite and the characteristics of microalgae after the experiment using (a) SEM, (b) XRD, (c) FTIR and (d, e) XPS. Fig. 6a shows that after photoreforming, brookite does not change in shape and size. Meanwhile, microalgae are significantly degraded after 3 h of irradiation, as the remaining particles are limited and their sizes are only 2–10 μm. XRD profiles (Fig. 6b) and XPS spectra of Ti 3p (Fig. 6d) before and after 3 h of photoreforming also confirm that the structure of brookite remains stable. It should be noted that Fig. 6c depicts the FTIR spectra of initial brookite, initial microalgae and the solid obtained after photoreforming. It is seen that after the reaction, the intensity of corresponding peak positions in the lipids and protein groups of microalgae declines, while the peak positions in the carbohydrate group disappear. This may indicate that the degradation ratio of microalgae from photocatalysis prioritizes carbohydrate transformation rather than degradation of complex functional groups such as lipids and proteins. Fig. 6e describes the XPS spectra of O 1s in the initial brookite and O 1s in the filtered solid after the reaction. Based on the peak deconvolution technique, it is shown that both spectra exhibit a peak at 529.8 eV, characteristic of lattice oxygen, with the peak position remaining constant after photocatalysis. Another peak appears at 531.0 eV, particularly after photocatalysis, which is characteristic of hydroxyl or oxygen defects [14] in brookite. These results indicate that while brookite remains stable during photocatalysis, microalgae are significantly degraded.

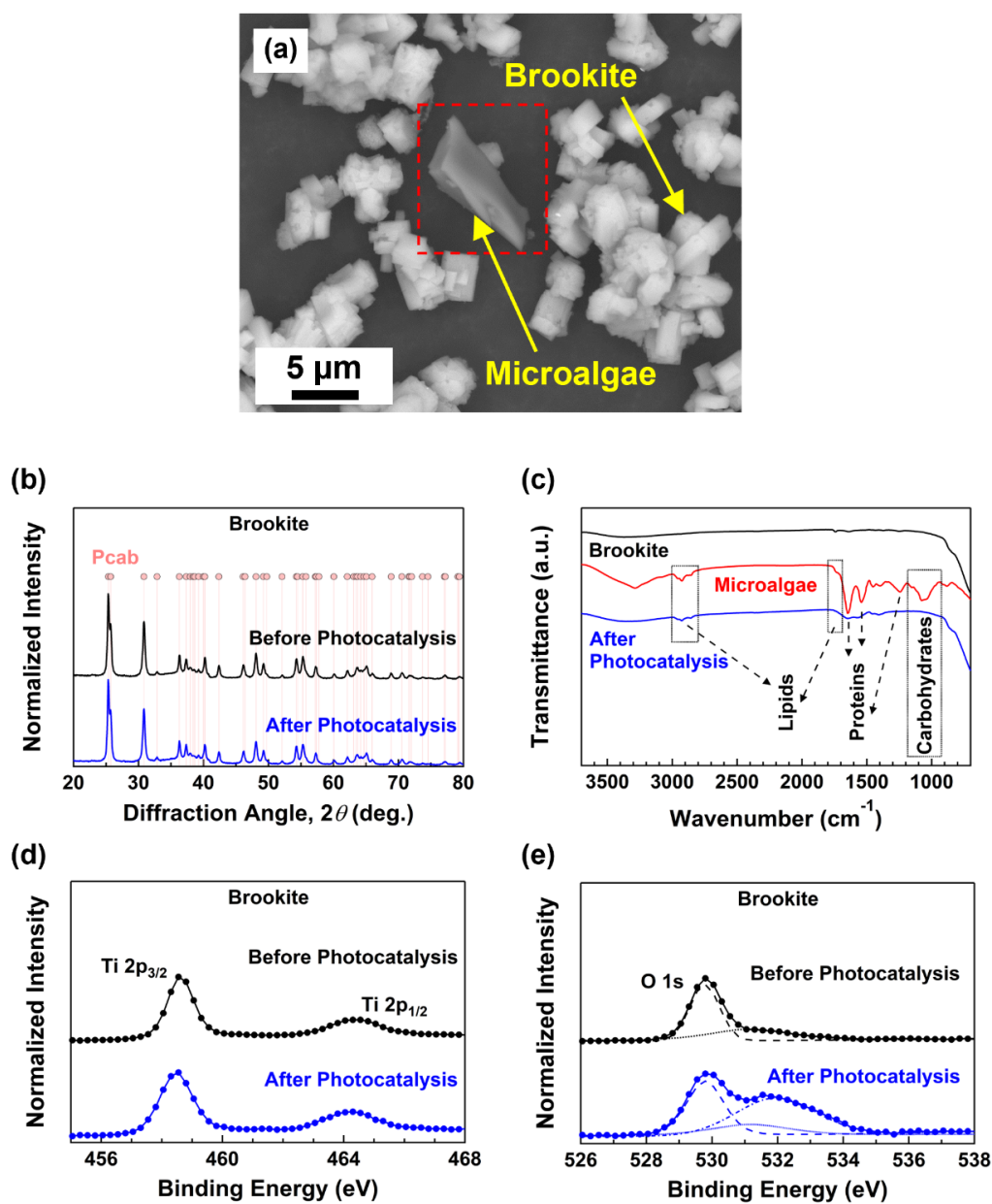

Fig. 6. Stability of brookite and degradation of microalgae after photocatalysis. (a) SEM image of filtered solid after 3 h photocatalysis, (b) XRD profiles of brookite before and after photocatalysis, (c) FTIR spectra of initial brookite, initial microalgae and filtered solid after photocatalysis, and (d) XPS Ti 2p and XPS O 1s spectra of brookite before and after 3 h photocatalysis.

## 4. Discussion

In this study, the application of microalgae as hole sacrificial agents in photocatalysis is introduced, which significantly improves the production efficiency of $H_2$ and leads to the production of other gases such as $CH_4$ and CO. Two issues should be discussed: (i) clarifying the role of microalgae in photocatalysis and describing the overall mechanism of the reaction, and (ii) the advantages of microalgae compared with other sacrificial agents.

To answer the first question, it is known for a long time that microalgae are an abundant and biodegradable source of biomass, with their main component including carbohydrates, proteins and lipids [26–29]. In a highly alkaline environment (10 M NaOH), microalgae are hydrolyzed and their structural chains break down into smaller components. As shown in the results from FT-IR analysis, the proposed products from the hydrolysis of microalgae in NaOH are identified as lipids, carbohydrates and proteins. Liquid-phase products such as carbohydrate, lipid and protein derivatives should be quantitatively analyzed using additional analytical procedures in future studies. When adding microalgae to the photocatalytic process, the fraction of $CH_4$, CO and particularly $H_2$ increases. This indicates that these intermediate components act as scavengers, combining with holes, preventing charge recombination and enhancing $H_2$ production from water, as reported in many previous studies [14,18], while their degradation also produces $CH_4$ and CO. Moreover, the disappearance of carbohydrates of microalgae also suggests that they convert to $CH_4$ and CO. The significance of microalgae as a sacrificial agent is evident from the comparison of the $H_2$ production in two photocatalytic reactions using brookite and 10 M NaOH with and without microalgae addition: 0.078 mmol/g.h without microalgae and 0.990 mmol/g.h with microalgae. In particular, under optimal conditions when adding platinum cocatalyst, the $H_2$ evolution efficiency can be up to 3.20 mmol/g.h in the presence of microalgae.

A proposed mechanism including two stages of microalgae photosynthesis and photoreforming is shown in Fig. 7. Microalgae play a central role in this process, as they can be cultivated using atmospheric $CO_2$ and photosynthesis to grow and develop while simultaneously releasing $O_2$. Once biomass reaches a certain level, the microalgae are harvested for photocatalytic applications. Although the carbon balance and exact conversion pathway cannot be commented on in the present study, it can be said that the dead microalgae, hydrolyzed in NaOH, react with photogenerated holes from brookite to form microalgae-derived intermediates, such as lipid, carbohydrate and protein compounds, and further byproducts, CO and $CH_4$. At the same time, the photocatalytic reduction half-reaction readily produces a large amount of H₂. In addition, platinum can be added during the microalgae photoreforming process to reduce the activation energy at the catalyst surface, leading to the highest energy evolution. It should be noted that since the catalyst maintains its structural integrity as described in Fig. 6b, d, e, the observed decline in cycling performance (Fig. 4d) is primarily due to depletion of the sacrificial microalgae rather than catalyst deactivation. The high stability and reusability of brookite for photocatalysis have also been widely reported in the literature [8-13], and thus, to maintain a consistently high yield, continuous addition of microalgae after each cycle is required.

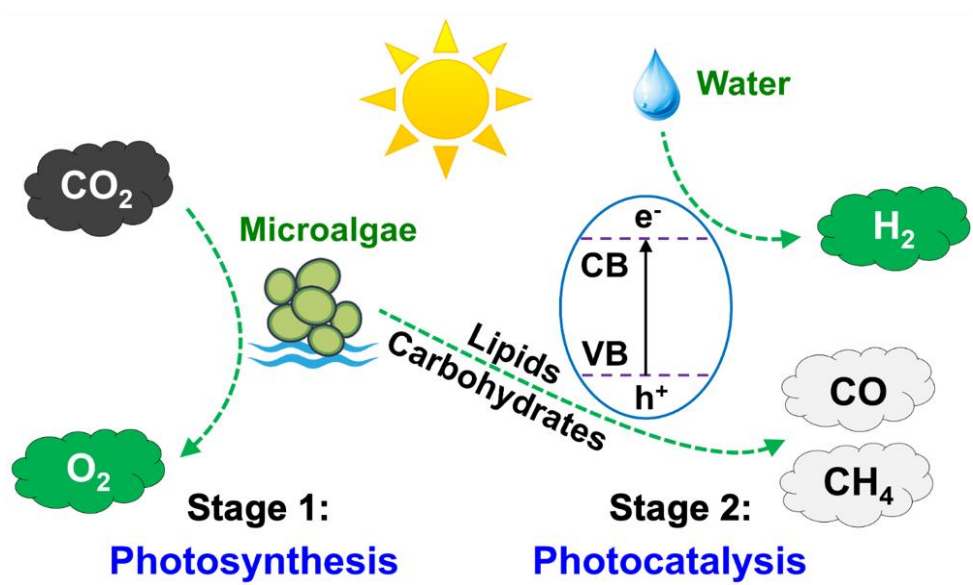

Fig. 7. Schematic illustration of simultaneous photocatalytic hydrogen production from water and microalgae photoconversion to valuable products such as CO and CH₄, in which microalgae also contribute to CO₂ capture via photosynthesis during their cultivation.

Regarding the second issue, we can refer to Table 1 to compare H₂ production from photocatalysis using different sacrificial agents [10.20,49-52]. Table 1 suggests that microalgae have a significantly higher potential than many other chemical sacrificial agents, although the comparison in Table 1 can be influenced by multiple factors such as reactor type, catalyst concentration, light intensity and energy, cocatalyst type, pH, etc. Chemical sacrificial agents, such as alcohol, require continuous production, which generates CO₂, while microalgae are renewable and can be cultivated using wastewater and capturing CO₂, aligning with environmental sustainability principles. Finally, beyond serving as electron donors, the intermediates from microalgae allow for more complex degradation and photoreforming pathways to produce biogases, such as $CH_4$ and CO. A modification of the process introduced in this study has potential for the generation of bioalcohol or biodiesel from microalgae.

Table 1. Summary of research on production of hydrogen by using different sacrificial agents under different photocatalytic conditions.

| Photocatalyst and Its Concentration | Catalytic Conditions | Scavenger for Holes | Hydrogen Production (mmol/g.h) | Other Products | Reference |
|---|---|---|---|---|---|
| Brookite (50 mg/ 27 mL) | - Solution: 27 mL NaOH 10 M, 250 μL $H_2PtCl_6·6H_2O$<br>- Lamp: Xenon (18 kW/m$^2$)<br>- Period: 3 h | Microalgae (50 mg/ 27 mL) | 3.20 | $CH_4$, CO | This Study |
| Brookite (50 mg/ 3 mL) | - Solution: 3 mL NaOH 10 M, 250 μL $H_2PtCl_6·6H_2O$<br>- Lamp: Xenon (18 kW/m$^2$)<br>- Period: 4 h | PET (50 mg/ 3 mL) | 2.27 | Acetic Acid | [10] |
| $CdO_x$/CdS/SiC (50 mg/ 5 mL) | - Solution: 5 mL NaOH 10 M,<br>- Pt 0.5 wt %-Deposited catalyst<br>- Lamp: Xenon (300 W) with Solar Simulator<br>- Period: 3 h | Organic waste (100 mg/ 5 mL)<br>- Cellulose<br>- Lignin<br>- Albumin<br>- Keratin | <br>0.36<br>0.07<br>0.65<br>0.45 | <br>- Glucose, 5-hydroxymethylfurfural and organic acids<br><br>- Amino Acids and Organic Acids | [20] |
| $TiO_2$ (100 mg/ 100 mL)<br>- Pt-Rutile<br>- Pd-Rutile<br>- Au-Rutile<br>- Pt-Brookite<br>- Pd-Brookite<br>- Au-Brookite | - Lamp: Xenon (200 W)<br>- Period: 5 h | Methanol 10 % (100 mL) | <br>- 1.90<br>- 1.60<br>- 0.70<br>- 1.50<br>- 1.30<br>- 0.60 | No Data | [49] |
| Bi-sphere-$C_3N_4$ (50 mg/ 100 mL) | - Solution: 100 mL Ultrapure Water<br>- Pt 1 wt %-Deposited Catalyst<br>- Lamp: Xenon (300 W) with a Cutoff Filter (λ > 420 nm)<br>- Period: 4 h | Amoxicillin (10 mg / 100 mL) | 0.72 | Amoxicillin Penicilloic Acid, Amoxicillin Penilloic Acid, Phenol Hydroxypyrazine, Amoxicillin 2',5'-Diketopiperazine, Amoxicillin-S-Oxide | [50] |
| CuO@exfoliated g–$C_3N_4$ (1 g/ 1 L) | - Lamp: LED Bulb (175 W/m$^2$)<br>- Period: 8 h | Formaldehyde (700 mg / 1 L) | 0.71 | No Data | [51] |
| Ni/Co$(PO_4)_2$-$TiO_2$ (3 mg/ 20 mL) | - Lamp: 1 Sun Conditions (100 mW/cm$^2$) with a 450 W Newport Solar Simulator Coupled with an AM 1.5 Filter<br>- Period: 3 h | Glycerol 5 % (20 mL) | 2.03 | Glyceraldehyde, Glyceric Acid, Dihydroxyacetone, CO | [52] |

Further studies should examine the electrochemical behaviour of photocatalysts in the presence of hydrolyzed microalgae using methods such as photocurrent and electrochemical impedance spectroscopy (EIS). Besides, the organic-phase extraction procedure for GC–MS analysis should be extended to include hydrophilic components of microalgae. Qualitative analyses, such as high-performance liquid chromatography (HPLC), should also be applied to assess mass balance and better understand the degradation mechanism of microalgae during photocatalysis. In the present experimental design, the use of 10 M NaOH is mitigated by the reuse of both the NaOH solution and the catalyst in continuous experiments, thereby reducing potential environmental impacts. However, microalgae that are effective under neutral pH conditions, together with different microalgal strains with natural nutrients or wastewater, together with other highly active catalysts, should be further investigated to enable scale-up, minimise environmental and energy concerns, and achieve maximum green fuel production. Although the yield of hydrogen production in this first attempt on the use of microalgae in a photocatalytic system is not yet comparable to dark fermentation [53], photobiological method [54] and electrolysis [55], significant improvement in this system is expected in the future. Since a wide range of active catalysts and processing modifications has been developed by the current authors and other scientists for photoreforming and/or hydrogen production [56-58], the authors believe an improvement in the yield should be feasible.

## 5. Conclusion

In this investigation, microalgae were used as green bio-sacrificial agents for photocatalytic $H_2$ generation. The hydrolysis of microalgae in an alkaline environment formed intermediate compounds, which acted as hole scavengers to enhance the efficiency of $H_2$ evolution. This new process also formed some other products, such as $CH_4$ and CO, which can be applied in industrial chemical processes. In addition to high efficiency and low cost, the main advantage of the introduced process is that microalgae capture and convert $CO_2$ during their cultivation, while the production of other sacrificial agents, like alcohols, generates $CO_2$, making their usage non-environmentally friendly. Therefore, harvesting microalgae through photosynthesis and its subsequent application to photocatalysis represents a sustainable strategy for reducing $CO_2$ while simultaneously enhancing energy product yields for not only $H_2$ production but also $CH_4$ and CO generation.

**CRediT authorship contribution statement**

All authors contributed to conceptualization, investigation, methodology, validation, and writing - review & editing.

**Declaration of competing interest**

The authors declare no conflict of interest.

**Data availability**

Data will be made available on request.

**Acknowledgments**

H.T.N.H. expresses sincere appreciation to the Yoshida Scholarship Foundation (YSF) for providing a Ph.D. scholarship. This research was supported in part by Mitsui Chemicals, Inc., Japan; in part by the ASPIRE project of the Japan Science and Technology Agency (JST) (JPMJAP2332); and in part by the São Paulo Research Foundation (FAPESP), grant number 2024/01639-0.